# Evolution of Mean Orbital Spacing in Planetary and Satellite Systems under Tidal Dissipation and Nebular Drag

Vladimir Pletser

Astronomy and Geophysics Institute G. Lemaitre, Catholic University of Louvain (ret.)
Louvain-la-Neuve, Belgium.
Email : Pletservladimir@gmail.com
ORCID : 0000-0003-4884-3827

## Abstract

The approximately geometric spacing of orbital distances in planetary and regular satellite systems has long been recognized, yet its dynamical evolution remains poorly constrained. In this paper, we investigate the secular evolution of the mean distance ratio of secondaries under the combined effects of primary tidal dissipation and nebular gas drag. A general analytical framework is derived linking the initial and final mean distance ratios to system parameters and to the physical characteristics of the dominant dissipative processes.

Applying this formalism to the Solar System planets and to the regular satellites of Jupiter, Saturn, and Uranus, we show that primary tidal interactions produce only negligible changes in mean distance ratios over timescales comparable to the age of the Solar System. Similarly, nebular gas drag during the protoplanetary and circumplanetary disk phases leads to limited deviations over a broad range of disk models and lifetimes.

These results suggest that, within the assumptions adopted, the mean distance ratio evolves only weakly and may preserve information about primordial system configurations established during early disk evolution. The approximate conservation of this quantity may therefore provide a useful diagnostic for constraining the formation and early dynamical evolution of planetary and satellite systems, with potential implications for the architecture of exoplanetary systems.

## 1. Introduction

The orbital architectures of planetary and satellite systems exhibit a number of well-known regularities, including low eccentricities, near-coplanarity, and the frequent occurrence of mean-motion resonances. In addition to these features, the approximate geometric spacing of orbital distances, often described through exponential or Titius–Bode–type relations, has long been recognized in the Solar System. However, the origin and long-term dynamical evolution of such spacing relations remain debated.

Modern theories of planet formation emphasize the key role of protoplanetary disks, within which planets form, migrate, and interact gravitationally with both gas and planetesimals. Disk–planet interactions can drive significant orbital migration, potentially reshaping planetary system architectures (e.g. Lin and Papaloizou 1986; Goldreich and Tremaine 1980). These processes are now understood to be central to the formation of both Solar System planets and extrasolar planetary systems (Ida and Lin 2004a, b; Morbidelli 2013; Youdin and Kenyon 2013).

Observations of exoplanetary systems have revealed a remarkable diversity of orbital architectures (Lissauer et al. 2011; Fabrycky et al. 2014), including compact multi-planet systems, resonant chains, and systems exhibiting quasi-regular spacing (e.g., Raymond et al. 2018, 2020; Bourrier et al. 2023). In particular, migration-driven evolution can produce systems with characteristic orbital spacing and compositional diversity (Raymond et al. 2018; Izidoro et al. 2022), while resonant capture and dynamical instabilities further modify orbital configurations (Batygin 2015; Barnes et al. 2015).

Despite this progress, it remains unclear to what extent large-scale structural properties, such as characteristic spacing ratios, are preserved during planetary system evolution. In particular, whether the approximately geometric spacing observed in planetary and satellite systems reflects primordial conditions or is significantly altered by subsequent dynamical processes is still an open question.

In this work, we investigate the evolution of the *mean distance ratio* of successive secondaries under the effects of primary tidal dissipation and nebular gas drag. Building on analytical formulations, we derive a general relation linking initial and final spacing properties to system parameters. We then apply this framework to the Solar System planets and to the regular satellites of the giant planets.

Our objective is to assess whether the observed spacing relations are robust against long-term dynamical evolution and to evaluate their potential as tracers of the physical processes governing planetary system formation. This question is particularly relevant in the context of modern models of planet formation and migration, which predict a wide range of possible orbital architectures (e.g., Raymond et al. 2022; Clement et al. 2024; McCloat et al. 2025).

## 2. Actual distance ratios

The semi-major axis of regular secondaries of the Solar System, the satellite systems of the giant planets, and of some exoplanetary systems are spaced in approximate geometrical progressions. This is expressed in exponential distance relations of the form

$$a_k = \alpha\, \beta^k \tag{1}$$

that are obeyed, with more or less accuracy, by all the planets and all the inner regular satellites of Jupiter, Saturn and Uranus for successive integers $k$ and where $a_k$ is the semi-major axis of the $k$-th secondary, $\alpha$ has dimensions of length and $\beta$ is a dimensionless spacing ratio, both being different for each system, with $1 < \beta < 2$. These relations, generalizing the Titius-Bode's planetary distance relation, fit well all the inner satellites distances, without discontinuity when one considers planetary rings and groups of small satellites (Pletser 1986, 1988a). Alternatively, the coefficients $\beta$ can be found approximately by geometrical means of the distance ratios of successive secondaries.

Such geometrical progressions have been debated for a long time. It was argued (Lecar 1973) that a similar relation for the planetary system can be obtained from distributions of random numbers and that planetary distances could be distributed at random under the constraint of a "not too close vicinity" of forming bodies. This was shown (Pletser 1987, 1988b, 2017a, b) to be a wrong conclusion, as the distribution of distances in the present planetary and satellite systems has more features than what could be accounted for by random distributions only. Alternatively, it can be argued that the observed near geometrical progressions between semi-major axes may result from cascading resonances or near-commensurabilities between secondaries' mean motions. Similar near geometrical progressions are observed in several exoplanetary systems (see e.g., Pletser and Basano 2017), reinforcing the hypothesis of their ubiquitous occurrence.

Table I shows the presently observed ratios $\beta_k$ of successive secondary semi-major axes $a_k$

$$\beta_k = \frac{a_k}{a_{k-1}} \tag{2}$$

for the planetary and satellite systems. Groups of small bodies with values of semi-major axis close to each other are represented by the largest body in the group: Ceres, Himalia, Pasiphae and Portia. Mean distances are considered for the rings (Pletser 1986, 1988a); however, distances for Jupiter's rings depend on ring models (Burns et al. 1984; Showalter et al. 1985) and their distance ratios in Table I are only indicative. Shepherd satellites and satellites associated with a ring are grouped with the ring.

The theoretical geometrical mean $\bar{\beta}$ of $n$ distance ratios $\beta_k$ of $(n+1)$ secondaries reads

$$\bar{\beta} = \sqrt[n]{\prod_{k=1}^{n} \beta_k} = \sqrt[n]{\frac{a_n}{a_0}} \tag{3}$$

Note that $\bar{\beta}$ depends only on the number of secondaries $n$ and the semi-major axes of the first and last secondaries considered. It means also that individual modifications of orbital distances of secondaries, other than the first and the last in the system, would not directly affect its value, provided that the semi-major axes of the first and last secondaries remain unchanged.

Table 1 indicates the observed geometrical means $< \beta >$ and associated standard errors of the distance ratios of the main regular secondaries, calculated without the outer irregular secondaries and without the inner rings and small satellites. These means are close to the ratios observed between rings and inner small satellites. For the Jovian system, the mean (1.596) is calculated over the set Amalthea to Callisto, counting a body between Amalthea and Io, as $a_{Io}/a_{Amalthea} = 2.325 = 1.525^2$. However, one should not be misled by the position of the small satellite Thebe ($\approx 40^3$ km$^3$) compared to the larger Amalthea ($\approx 135\times85\times75$ km$^3$) and to the Galilean satellites. For Saturn's system, the mean ratios 1.458 (Mimas to Titan) and 1.525 (Mimas to Iapetus) include one and two large gaps (Rhea/Titan and Hyperion/Iapetus), as attested by the larger standard errors, and do not represent most of the distance ratios of the rest of the inner system. A mean ratio 1.298 (Mimas to Rhea) better represents these distance ratios and is also closer to the ratios observed among the satellites and rings inside Mimas' orbit. Gaps in the Saturn satellites sequence were already discussed (Pletser 1986). Note that, while the two interlocked pairs Mimas/Tethys and Enceladus/Dione are in resonances 4/2 and 2/1 (see e.g., Greenberg 1984), there are no known independent resonances for the pairs Mimas/Enceladus, Enceladus/Tethys or Tethys/Dione. But one notices the remarkable closeness of the ratios of successive distances between the four satellites.

Table 1: Distance ratios in actual planetary and satellite systems

| Planetary | $\beta_k$ | Jupiter | $\beta_k$ | Saturn | $\beta_k$ | Uranus | $\beta_k$ |
|---|---|---|---|---|---|---|---|
| Mercury | | (Halo ring) | | (Ring D) | | (Rings) | |
| Venus | 1.868 | (Main ring) | 1.167 | (Ring C) | 1.196 | (Portia) | 1.560 |
| Earth | 1.383 | Amalthea | 1.438 | (Ring B) | 1.261 | Puck | 1.378 |
| Mars | 1.524 | (Thebe) | 1.224 | (Ring A) | 1.254 | Miranda | 1.511 |
| Ceres | 1.816 | Io | 1.900 | (Janus) | 1.146 | Ariel | 1.470 |
| Jupiter | 1.880 | Europe | 1.591 | Mimas | 1.225 | Umbriel | 1.393 |
| Saturn | 1.837 | Ganymede | 1.595 | Enceladus | 1.283 | Titania | 1.640 |
| Uranus | 2.011 | Callisto | 1.757 | Tethys | 1.238 | Oberon | 1.337 |
| Neptune | 1.567 | [Himalia] | 6.101 | Dione | 1.281 | | |
| [Pluto] | 1.310 | [Pasiphae] | 2.031 | Rhea | 1.397 | | |
| | | | | Titan | 2.318 | | |
| | | | | Hyperion | 1.212 | | |
| | | | | Iapetus | 2.404 | | |
| | | | | [Phoebe] | 3.638 | | |
| $<\beta>_{Me/Ne}=$ $1.723 \pm 0.072$ | | $<\beta>_{Am/Ca}=$ $1.596 \pm 0.055$ | | $<\beta>_{Mi/Ti}=$ $1.458 \pm 0.185$ | | $<\beta>_{Mi/Ob}=$ $1.456 \pm 0.057$ | |
| | | | | $<\beta>_{Mi/Ia}=$ $1.525 \pm 0.187$ | | | |
| | | | | $<\beta>_{Mi/Rh}=$ $1.298 \pm 0.029$ | | | |

$\beta_k = \frac{a_k}{a_{k-1}}$ : ratio of semi-major axes of successive secondaries

$<\beta> \pm \frac{s}{\sqrt{n}}$ : geometrical mean and associated standard error, for n ratios of main regular secondaries.

Parentheses indicate small inner secondaries and rings; brackets indicate outer irregular secondaries.

### 3. Evolution of a mean distance ratio

Considering the theoretical mean distance ratio (3) as a characteristic parameter of the system, its time derivative (denoted by an upper dot) reads

$$\dot{\bar{\beta}} = \frac{1}{n} \sqrt[n]{\prod_{k=1}^{n} \beta_k} \sum_{k=1}^{n} \frac{\dot{\beta}_k}{\beta_k} = \frac{\bar{\beta}}{n} \sum_{k=1}^{n} \left( \frac{\dot{a}_k}{a_k} - \frac{\dot{a}_{k-1}}{a_{k-1}} \right) = \frac{\bar{\beta}}{n} \left( \frac{\dot{a}_n}{a_n} - \frac{\dot{a}_0}{a_0} \right) \tag{4}$$

It depends on the mean distance ratio itself, on the number of ratios in the mean and on the relative changes in the semi-major axis of only the first and the last secondaries considered in the mean.

This formulation highlights that the evolution of the mean distance ratio depends primarily on boundary conditions within the system, rather than on the detailed evolution of each individual secondary, making it a robust global descriptor of system architecture.

The relative change in a secondary $k$ semi-major axis due to both the primary tidal effect and the gas drag effect can be similarly written in general as

$$\frac{\dot{a}_k}{a_k} = C^* \, C_k \, a_k^q \tag{5}$$

where $C^*$, $C_k$ and $q$ are parameters depending on the kind of dynamical effect, $C^*$ represents system-dependent parameters, i.e. for the whole system, including the central primary, and $C_k$ accounts for body-specific properties such as mass or radius of the $k$-th secondary.

Replacing in (4) yields successively

$$\dot{\bar{\beta}} = \frac{\bar{\beta}}{n} \frac{\dot{a}_0}{a_0} \left( \frac{\dot{a}_n}{a_n} \frac{a_0}{\dot{a}_0} - 1 \right) = \frac{\bar{\beta}}{n} \frac{\dot{a}_0}{a_0} \left( \frac{C_n}{C_0} \left( \frac{a_n}{a_0} \right)^q - 1 \right) \tag{6}$$

$$\frac{\dot{\bar{\beta}}}{\bar{\beta}} \frac{1}{\left( \frac{C_n}{C_0} \, \bar{\beta}^{nq} - 1 \right)} = \frac{1}{n} \frac{\dot{a}_0}{a_0} \tag{7}$$

Assuming that $C_n$ and $C_0$ are independent of time $t$, the integration of (7) with respect to time between initial and final epochs $t_i$ and $t_f$ gives

$$\left( \frac{C_n}{C_0} - \frac{1}{\bar{\beta}_i^{nq}} \right) \frac{1}{a_{0.i}^q} = \left( \frac{C_n}{C_0} - \frac{1}{\bar{\beta}_f^{nq}} \right) \frac{1}{a_{0.f}^q} \tag{8}$$

where $\bar{\beta}_i = \bar{\beta}(t_i)$ and $a_{0.i} = a_0(t_i)$ are the mean distance ratio and the first secondary semi-major axis at the initial epoch, and similarly for the final epoch. Assuming that $C^*$ is also independent of time, the integration of (5) between $t_i$ and $t_f$ gives

$$\left(\frac{a_{0.i}}{a_{0.f}}\right)^q = \frac{1}{1 + qC^* C_0 a_{0.f}^q (t_f - t_i)} = \frac{1}{1+K}$$

(9)

Replacing in (8) yields an estimation of the mean distance ratio at the initial epoch as a function of the mean distance ratio and of characteristics of the primary and the first secondary at the final epoch, or

$$\frac{\bar{\beta}_i}{\bar{\beta}_f} = \sqrt[nq]{\frac{1+K}{1+K\,\frac{C_n}{C_0}\,\bar{\beta}_f^{nq}}} = 1 + \Delta$$

(10)

where $\Delta$ is the deviation of the ratio $\frac{\bar{\beta}_i}{\bar{\beta}_f}$ from unity. If $K$ is negative (e.g., the exponent $q$ or terms in $C^*$ or $C_0$ could be negative), then obviously the term under the $nq$-root in (10) has to be positive for this method to be valid, restricting the choice of the first or last secondaries over which the mean is calculated. For large systems, i.e. large $n$, the $nq$-root smooths out any change out of proportion with the mean, in individual distance ratios of successive secondaries evolving differently from other secondaries.

**4. Models and hypotheses**

The effects of primary tides and nebular drag on the secondaries' mean distance ratios are estimated using the above method, where calculations are conducted on orders of magnitude. Two stages are considered: the initial nebular stage, where proto-secondaries formed and evolved in a gaseous environment, and a second stage, starting after the dissipation of the initial nebulae and lasting till nowadays.

At the quasistatic transition, when the proto-primaries passed from a rapid hydrodynamic collapse phase to a quasi-hydrostatic contraction phase, the proto-Sun and the giant proto-planets are assumed to have reached approximately their present masses, while their radii were larger than presently, about 50 to 60, 1.3 and 3.4 times the present values of respectively the Sun (see e.g., Cox and Giuli 1968), Jupiter and Saturn (Bodenheimer et al. 1980). Initial decrease of the radii to approximately their present values took place within roughly the first few $10^7$ yrs for the proto-Sun (Iben 1965) and for proto-Jupiter (Bodenheimer et al. 1980; Grasboke et al. 1975) and within the first few $10^5$ to $10^6$ yrs for proto-Saturn (Pollack et al.

1977), and most probably during the nebular stage. An initial radius of 2 $R_U$ is assumed for proto-Uranus, and its initial decrease is assumed short as for Jupiter and Saturn.

Two nebula models are usually considered: the low and high mass models, with a nebula mass respectively small in front of the central primary mass (typically about a few percent) or of the order of the central mass. However, in the planetary high mass model (Cameron and Pine 1973), about 10% of the mass is in the part of the disc within the present distance of Neptune (Stevenson et al. 1986), so in that respect this model is only a few times more massive than the low mass model.

The planetary nebula lifetime in the low mass case is generally assumed to be in the order of $10^6$ (Weidenschilling 1977a) to $10^7$ yrs (Lin 1981; Hayashi et al. 1985), similar to the T Tauri stage of stellar evolution (Ezer and Cameron 1965), while for the high mass model, the lifetime is in the order of $10^4$ yrs. Lifetimes of satellites nebulae depend mostly on hypotheses on supply of heliocentric nebular material (gas and solid particles or planetesimals) to the circumplanetary nebulae: either the supply is continuous by infalling of heliocentric material onto the circumplanetary nebulae, yielding continuous accretion of planetocentric nebular material onto the giant proto-planets by inward spiralling (Harris 1978, 1984), in which case the nebulae lifetimes are in the order of about $10^6$ yrs or less (Pollack 1985); or this supply is not possible because the giant proto-planets tidally truncated the Solar nebula in their vicinity (Lin and Papaloizou 1980), in which case the lifetimes are much shorter.

Drag on secondaries is caused mainly by the nebular gas but also by impacting solid particles and planetesimals, the former being active only during the nebular stage, while the latter may also exist after the nebulae dissipations (negligible drag is presently due to meteoritic bombardment). Additionally, forming primaries accrete nebular material by inward spiralling, causing secondaries' orbits to decay and secondly, forming secondaries may also accrete gas and impacting solid particles, this "accretion drag" causing also orbits to decay (Harris 1978; Burns 1977). However, the effect of accreting primaries may be less important toward the end of the nebular stage and gas accretion by secondaries requires adequate mass and temperature conditions, while impacting planetesimals may also cause loss of secondaries' material, depending on the impactors' velocity and mass. On the other hand, the nebular gas specific mass may be locally decreased at distances of forming secondaries as gas may be accreted by secondaries or because of the tidal tunnelling effect caused by the secondaries in the nebula: the gas is tidally "pushed away" on either side of the secondary orbit (similarly to shepherd satellites' effect on planetary rings) when the secondary reaches a certain critical mass (Lin and Papaloizou 1979: Lynden-Bell and PringIe 1974: Coradini et al. 1981). This tidal effect clears a "tunnel” of low gaseous specific mass in the nebula, reducing the accretion and diminishing the drag (Weidenschilling 1982). It reduces the period during which the nebular drag is effective to the period of growth of the proto-secondaries, up to reaching the minimum mass to trigger the tunnelling effect. These minimum masses in the Jovian and Saturnian nebulae are in the order of $10^{21}$and $10^{18}$ kg (Weidenschilling 1982), which are respectively about 50 to 150 times less than the present masses of the Galilean satellites and about 40, $2.4\times10^3$ and $1.3\times10^5$ times less than the present masses of respectively Mimas, Rhea and Titan. For comparison, the timescales of accretion of solid cores of the Galilean satellites range between several $10^2$ to several $10^4$ yrs (Weidenschilling 1982). Therefore, the tidal tunnelling effect may have been triggered quite early in the nebular stage. However, conversely, this tidal tunnelling effect could not have been stable for long, as turbulence and initial crossing orbits may destroy or smooth out gas cleared tunnels (Safronov et al. 1986).

In light of these considerations, we make the following hypotheses and assumptions, such as to give upper limits of the tidal and drag effects on the evolution of the secondaries' mean distance ratios, although some crude assumptions are made for the gas drag effect to grossly balance the conflicting uncertainties on nebular gas specific mass, body sizes and duration of effective drag.

1) The position orders of the present main regular planets and satellites are assumed primordial; no past drastic exchanges of orbits of regular planets and satellites are envisaged (however, see further).

2) Although drag and tidal effects have been acting simultaneously since the formation of the systems, it is considered that the drag was predominant over the primary tidal effect during the nebular stage (Harris 1978), while after the dissipation of the nebulae, only the tidal effect prevails (Burns 1977, 1982).

3) For the primary tidal effect, present values are considered for the primary radii and for the masses of primaries and secondaries, as it is assumed that the decrease of proto-primaries radii takes place mostly during the nebular stage and that variations in main secondaries' masses are small after the nebulae dissipations.

4) The period $(t_f - t_i)$ for secondaries' tidal evolution is taken to be $4.5\times10^9$ yrs, as the nebular stages are short on astronomical scales and most probably less than 1% of the Solar System age.

5) Drag is considered to be caused by the gas circular velocity only; drag caused by the radial and vertical components of the gas velocity and drag due to impacting particles or planetesimals are neglected.
6) Initial radial distributions of gas specific mass, temperature and sound speed in the nebulae are supposed to be continuous and decreasing outwardly from central regions. Nebulae are considered to retain their initial distributions of temperature and sound speed, although initial values were probably larger than later values, decreasing with time during the nebular stage.
7) Primaries during the nebular stage are considered with their present masses but with their larger initial radii at the beginning of the quasi-hydrostatic contraction.
8) Non-accreting secondaries are considered with their present masses and dimensions, revolving on coplanar circular orbits, as gas drag tends to circularise orbits (Burns 1976).
9) Due to the tidal tunnelling effect, the duration of effective drag may be less than the nebulae's lifetimes, and the specific mass of the gas effectively encountered by the secondaries may be less than what would be inferred from a continuous distribution.
These assumptions are consistent with current models of planet formation, in which bodies form within evolving protoplanetary disks through a combination of accretion and migration processes (Youdin and Shu 2002; Youdin and Kenyon 2013). However, the first hypothesis is somehow contradicted by results of numerical simulations. According to some of the Nice model simulations (Tsiganis et al. 2005; Desch 2007; Morbidelli et al. 2007, 2012), the initial orbit of proto-Neptune started inside the orbit of proto-Uranus; both proto-giant planets exchanged places early during the Solar System formation. However, as pointed by several authors including the Nice model proponents (Raymond et al. 2022), there is as-yet no consensus model for the Solar System formation. One can then assume that the first hypothesis is plausible.

## 5. Expression of $\frac{\bar{\beta}_i}{\bar{\beta}_f}$ for the primary tidal effect

The tides caused by secondaries on their primary cause the secondaries' semi-major axis to decrease or increase, depending on whether the secondary revolves inside or outside the primary synchronous orbit.
The relative rate of change of a secondary $k$ semi-major axis $a_k$ due to primary tides reads (Burns 1977)

$$\frac{\dot{a}_k}{a_k} = \frac{\pm 3\sqrt{\frac{G}{M^*}}\, R^{*5}\, k_2}{Q}\, m_k\, a_k^{\frac{-13}{2}} = C^*\, C_k\, a_k^q$$

(11)

where the positive (or negative) sign is taken for an outward (or inward) evolution of the $k$-th secondary and with $G$ the gravitational constant, $M^*$ the primary mass, $R^*$ the primary radius, $k_2$ the potential Love number, $Q$ the tidal specific dissipation factor and $m_k$ the secondary mass. At similar distances, massive secondaries will evolve faster, while for similar masses, secondaries closer to the primary will evolve faster. As $q = -\frac{13}{2}$ is negative and assuming the positive sign of $C^*$ in (11), the ratio of initial to final mean distance ratios is

$$\frac{\bar{\beta}_i}{\bar{\beta}_f} = \sqrt[\frac{13n}{2}]{\frac{1 + K_{Tide}\, \frac{m_n}{m_0}\, \bar{\beta}_f^{\frac{-13n}{2}}}{1 + K_{Tide}}} = 1 + \Delta_{Tide}$$

(12)

with

$$K_{Tide} = -\frac{39}{2}\, \frac{\sqrt{\frac{G}{M^*}}\, R^{*5}\, k_2}{Q}\, m_0\, \frac{1}{\sqrt{a_{0.f}^{13}}}\left(t_f - t_i\right)$$

(13)

within the hypotheses that $M^*$, $R^*$, $k_2$, $Q$, $m_0$ and $n$ are independent of time during the period $\left(t_f - t_i\right)$.

## 6. Evolution of $\frac{\bar{\beta}_i}{\bar{\beta}_f}$ for the four systems under primary tidal effect

The relation (12) is applied to the four systems, even if the method of tidal evolution (11) may not be applied separately to secondaries in resonance, as in the Jovian and Saturnian systems (see further).

The secondary position relative to the primary-synchronous orbit determines its inward or outward evolution. The radial distance of the synchronous orbit for a primary spinning at an angular velocity $\omega^*$ is

$$a_{sync} = \sqrt[3]{\frac{GM^*}{\omega^{*2}}}$$

(14)

The Sun is generally recognised to have rotated initially faster than presently, losing its angular momentum either by magnetic or viscous transfer to outer parts of the planetary nebula or by ejection of strong stellar wind (Hayashi et al. 1985) and most probably during or at the end of the nebular stage. For a faster rotation, the Sun's $a_{sync}$ would have been smaller than presently and would have progressively increased due to spin decay. As the present $a_{sync}$ is about 0.17 AU, halfway to Mercury, the planets after the nebular stage are assumed to always have been beyond the Sun-synchronous orbit and to have tidally evolved outward.

As the rotation of giant planets is scarcely affected by tides caused by the satellites, one assumes the conservation of the rotation angular momentum after the giant planets' accretion, to give an upper limit estimate on their synchronous orbit locations after the nebular stage. To account for the total mechanical energy of a planet (gravitational potential plus rotational and orbital kinetic energies, considering the losses for the planet's interior heating and outward radiation) or to consider a faster initial rotation would yield a higher initial value of $\omega^*$, bringing the synchronous orbit closer to the planet. The present $a_{sync}$ of Jupiter, Saturn and Uranus are respectively 2.24 $R_J$, 1.86 $R_S$ and 3.26 $R_U$, inside respectively the present orbits of Amalthea, the B ring and Puck. Initial radii of 1.3 $R_J$, 3.4 $R_S$ and 2 $R_U$ (assumed) yield initial $a_{sync}$ values of 3.18 $R_J$, 9.6 $R_S$ and 8.2 $R_U$, respectively beyond Amalthea's, Rhea's and Ariel's present orbits. We assume then that the Galilean satellites, Titan, Umbriel and further Saturnian and Uranian satellites were always beyond the synchronous orbit after the nebular stage, while Amalthea, Rhea, possibly Ariel and inner satellites evolved first inward and then outward when they crossed the recessing synchronous orbit due to the planet's radius decrease. However, the timescale of this inward-outward mechanism is short, as Jupiter's radius decreased in the first few $10^7$ yrs or so and as Saturn's radius decreased from 3.4 to 1.5 $R_S$ within only $5\times10^5$ yrs (Pollack et al. 1977). As the timescales of recession of proto-primaries radii and synchronous orbits are about the lifetimes of the nebular stages, an outward tidal evolution after the nebulae dissipations can then be assumed for all the main secondaries (excluding, of course, the outer retrograde irregular satellites).

The Love number $k_2$ ranges between theoretical values 0 and 1.5 for bodies of infinite and null rigidities (perfect solid and perfect fluid). Values about 0.4 are usually found for fluid primaries (Burns 1977). Calculations of tidal effects on giant planets yield values of 0.379, 0.341 and 0.104, respectively for Jupiter, Saturn and Uranus (Gavrilov and Zharkov 1977).

The values of the tidal dissipation factors $Q$ are more debated. For the planetary system, instead of the Sun's $Q$, we consider the theoretical maximum value of $Q^{-1} \approx \sin 2\varepsilon \approx 1$, where $\varepsilon$ is the tidal lag angle. Lower bounds on $Q$ for Jupiter, Saturn and Uranus were given by Goldreich and Soter (1966), respectively $10^5$, $6\times10^4$ and $7\times10^4$, using a value $k_2 = 1.5$ of pure fluid. With the $k_2$ values of Gavrilov and Zharkov (1977), these bounds are $2.5\times10^4$, $1.4\times10^4$ and $5\times10^3$ (Schubert et al. 1986).

Results of calculations on magnitude order are given in Table 2 for several combinations, with upper limits on $Q^{-1}$, and with $< \beta_f >$ taken from Table 1 as estimates for $\bar{\beta}_f$. All the deviations $\Delta_{Tide}$ of $\frac{\bar{\beta}_i}{\bar{\beta}_f}$ from unity are small and positive, showing that the mean distance ratios did not evolve significantly over $4.5\times10^9$ yrs under primary tidal actions only, and that the initial mean distance ratios were probably a little larger than presently. Recall that this result does not imply that neither the distances nor individual distance ratios are conserved. Most remarkably for the planetary systems, even with the theoretical upper limit $Q^{-1} = 1$, the ratio $\frac{\bar{\beta}_i}{\bar{\beta}_f}$ is close to unity within $7.3\times10^{-7}$, $2\times10^{-6}$ and $5.2\times10^{-10}$, respectively for the sequence Mercury to Neptune (counting a body in the Asteroid belt), for the four terrestrial planets and for the four giant planets.

Table 2: Deviation $\Delta_{Tide} = \frac{\bar{\beta}_i}{\bar{\beta}_f} - 1$ in the four systems under primary tidal effect

| System | $n$ | Secondaries set | $< \beta_f >$ | $k_2$ | $Q$ | $\Delta_{Tide}$ |
|---|---|---|---|---|---|---|
| Planetary | 8 | Mercury-Neptune ([a]) | 1.723 | 0.4 | 1 ([c]) | $7.3\times10^{-7}$ |
| | 3 | Mercury-Mars | 1.579 | 0.4 | 1 ([c]) | $2.0\times10^{-6}$ |
| | 3 | Jupiter-Neptune | 1.796 | 0.4 | 1 ([c]) | $5.2\times10^{-10}$ |
| Jupiter | 5 | Amalthea-Callisto ([b]) | 1.596 | 0.379 | $10^5$ ([d]) | $1.8\times10^{-3}$ |
| | 5 | Amalthea-Callisto ([b]) | 1.596 | 0.379 | $2.5\times10^4$ ([e]) | $7.7\times10^{-3}$ |
| Saturn | 4 | Mimas-Rhea | 1.298 | 0.341 | $6\times10^4$ ([f]) | $1.3\times10^{-2}$ |
| | 5 | Mimas-Titan | 1.458 | 0.341 | $6\times10^4$ ([f]) | $1.1\times10^{-2}$ |
| | 7 | Mimas-Iapetus | 1.525 | 0.341 | $6\times10^4$ ([f]) | $8.2\times10^{-3}$ |
| | 4 | Dione-Iapetus | 1.752 | 0.341 | $1.4\times10^4$ ([e]) | $1.6\times10^{-2}$ |
| Uranus | 4 | Miranda-Oberon | 1.456 | 0.4 | $7\times10^4$ ([d]) | $9.0\times10^{-3}$ |
| | 4 | Miranda-Oberon | 1.456 | 0.104 | $7\times10^4$ ([d]) | $2.1\times10^{-3}$ |
| | 4 | Miranda-Oberon | 1.456 | 0.104 | $5\times10^3$ ([e]) | $5.6\times10^{-2}$ |

(a) including one body in the Asteroid belt

(b) including one body between Amalthea and Io

(c) theoretical maximum value of $Q^{-1} \approx \sin 2\varepsilon \approx 1$

Upper limit value on $Q^{-1}$ from (d) Goldreich and Soter (1966), (e) Schubert et al. (1986); (f) Goldreich and Soter (1966) and Yoder (1981).

For the Jovian and Uranian systems, the deviations are less than 1%, except for one case in the Uranian system (about 5%) where a low value of $Q$ was used. For Saturn's system, the deviations for the combinations presented are all less than 1.6%. However, the values of $Q$ for the giant planets are still debated and depend on models (see e.g. Schubert et al. 1986). Note that higher values of $Q$ would diminish the deviations $\Delta_{Tide}$ and render the ratios $\frac{\bar{\beta}_i}{\bar{\beta}_f}$ even closer to unity.

As an inner satellite evolves outwardly faster than satellites further away from the primary, due to the term $a_k^{\frac{-13}{2}}$ in (11), it may approach a slower evolving satellite, such as the ratio of their mean motions comes close to a small integer fraction, and both satellites may possibly enter in resonance. Such mutual gravitational interactions between resonant satellites are known to be strong enough to resist further separate tidal evolution (Goldreich 1965). The pair evolves then outwardly together as primary tides drive the innermost satellite outward, which in turn pushes away the satellite with which it is in resonance (Burns 1977). The Laplace resonance relation between Io, Europe and Ganymede was considered unlikely to have been caused by tidal evolution only (Sinclair 1975) and to be probably primordial (Greenberg 1982). However, it was shown that the Laplace three-body relation can be explained in terms of two bodies' tidal capture, accounting for the dissipation of tidal energy in Io, giving an upper bound on $Q_J$ of $3\times10^6$, and leading to present equilibrium configurations of the three satellites (Yoder 1979; Yoder and Peale 1981; Henrard 1983: Peale 1986). On the other hand, tidal episodic evolutions out of deep resonance of the three Galilean satellites are discussed in (Greenberg et al.,1986; Greenberg 1987).

In any case, the distance ratio of satellites locked in resonance does not change secularly and remains the same since the epoch of the locking in resonance. Such local freezing of distance ratios between satellites does not influence the global evolution of the system mean distance ratio, but reinforces a fortiori its approximate conservation. The method of tidal evolution and the relation (11) are not applicable separately to satellites once they are locked in resonance.

However, as the ratio $\frac{\bar{\beta}_i}{\bar{\beta}_f}$ depends on characteristics of only the first and last secondaries considered in the mean distance ratio and provided that these do not form a resonant pair, the results for the Jovian and Saturnian systems in Table 2 are significant.

From a contemporary perspective, this result is consistent with the view that tidal dissipation primarily affects inner system bodies and short-range interactions, while leaving large-scale structural properties of planetary systems largely unchanged.

## 7. Expression of $\frac{\bar{\beta}_i}{\bar{\beta}_f}$ for the gas drag effect

A body orbiting in a differentially rotating gaseous nebula experiences a drag caused by the difference between the gas' circular velocity and the body's orbital Keplerian velocity, as the gas is supported by its pressure gradient. For a negative pressure gradient, the gas velocity is smaller than the body's velocity, and the drag causes the body's orbit to decay. The expression of the drag force depends on the gas flow regime and on the mean free path $\lambda_g$ of the gas molecules, being smaller or larger than a body's characteristic distance $L$.

As we look at secondaries of relatively large sizes moving in gaseous nebulae, the flow regime is characterised by large Reynolds numbers ($Re$). Stokes' or Epstein's drag laws are therefore not considered, the former being applicable for small $Re$ and the latter for cases where $\lambda_g \geq L$.

In general, the drag force acting on a body moving at a velocity $v$ relative to a gas of specific mass $\rho$ reads then

$$F = \frac{1}{2} C_D \, \rho \, S \, v^2 \tag{15}$$

where $S$ is the body cross-section area normal to $v$ and $C_D$ is a dimensionless parameter depending on the body geometry and the flow regime. Assuming secondaries of spherical shape of radius $R_k$, $C_D \approx 0{,}44$ for $Re > 800$ (Whipple 1972). The relative velocity of the $k$-th secondary is

$$v = v_K - v_g = \sqrt{\frac{GM^*}{a_k}} - \sqrt{\frac{GM^*}{a_k} + \frac{a_k}{\rho}\frac{\partial p}{\partial r}} \tag{16}$$

where the gas pressure gradient $\frac{\partial p}{\partial r}$ and specific mass $\rho$ have to be evaluated at the distance $r = a_k$. In planetary and satellite nebulae, the part of the gas circular velocity due to the pressure gradient is much smaller than the one due to the gravitational potential (Weidenschilling 1977a), yielding

$$\frac{a_k^2}{GM^*}\frac{\partial p}{\partial r}\frac{1}{\rho} \ll 1 \tag{17}$$

and the relative velocity reduces to

$$v \approx -\frac{1}{2}\sqrt{\frac{a_k^3}{GM^*}\frac{\partial p}{\partial r}\frac{1}{\rho}} \tag{18}$$

The pressure is expressed as a function of the sound speed $c$. The gas specific mass and sound speed are expressed as power-law distributions in the radial distance $r$ in the nebula midplane.

$$p = c^2\,\frac{\rho}{\gamma} \quad ; \quad \rho = \rho_c\left(\frac{r}{r_c}\right)^d \quad ; \quad c = c_c\left(\frac{r}{r_c}\right)^{\frac{s}{2}} \tag{19}$$

where $\gamma$ is the polytropic index, $d$ and $s$ are model-dependent exponents, but usually negative and $\rho_c$ and $c_c$ are values of the gas specific mass and sound speed at a reference distance $r_c$. The exponent $s$ is approximated in the perfect gas hypothesis by the exponent of the temperature radial distribution, of values $-\frac{1}{2}$, $-1$ or about $-\frac{3}{2}$ for respectively optically thin, optically thick or viscous nebulae (Pollack and Consolmagno 1984; Lin 1981).

The relative velocity (18), with (19), becomes

$$v \approx -\frac{(s+d)}{2}\sqrt{\frac{1}{GM^*}\frac{c_c^2}{\gamma\, r_c^s}}\; a_k^{s+\frac{1}{2}} \tag{20}$$

and is independent of the radial distance for an optically thin nebula ($s = -\frac{1}{2}$).

The drag force acting on a secondary $k$ at a distance $a_k$ and caused by the nebular gas reads then

$$F = \frac{\pi}{8} C_D\,(s+d)^2\,\frac{c_c^4}{\gamma^2}\frac{\rho_c}{r_c^{2s+d}}\,\frac{R_k^2}{GM^*}\,a_k^{2s+d+1} \tag{21}$$

The rate of change of the semi-major axis $a_k$ of a secondary of mass $m_k$ on a circular orbit due to a force $F$ tangential to the orbit generally reads (Burns 1976)

$$\dot{a}_k = 2\sqrt{\frac{a_k^3}{GM^*}}\frac{F}{m_k}$$

(22)

Replacing $F$ with (21) yields the relative rate of change in $a_k$ due to gas drag

$$\frac{\dot{a}_k}{a_k} = \left(-\frac{\pi}{4}C_D\,(s+d)^2\frac{c_c^4}{\gamma^2}\frac{\rho_c}{r_c^{2s+d}}\frac{1}{\sqrt{(GM^*)^3}}\right)\left(\frac{R_k^2}{m_k}\right)a_k^{2s+d+\frac{3}{2}} = C^*C_k a_k^q$$

(23)

where the negative sign arises because the gas drag causes $a_k$ to decrease. This relation is compared to (5), giving the expressions of $C^*$, $C_k$ and $q$. The ratio $\frac{\bar{\beta}_i}{\bar{\beta}_f}$ reads then from (10)

$$\frac{\bar{\beta}_i}{\bar{\beta}_f} = \sqrt[n\left(2s+d+\frac{3}{2}\right)]{\frac{1+K_{Drag}}{1+K_{Drag}\left(\frac{R_n}{R_0}\right)^2\frac{m_0}{m_n}\,\bar{\beta}_f^{\,n\left(2s+d+\frac{3}{2}\right)}}} = 1+\Delta_{Drag}$$

(24)

with

$$K_{Drag} = -\frac{\pi}{4}C_D\left(2s+d+\frac{3}{2}\right)(s+d)^2\frac{c_c^4}{\gamma^2}\frac{\rho_c}{r_c^{2s+d}}\frac{1}{\sqrt{(GM^*)^3}}\frac{R_0^2}{m_0}a_{0.f}^{\;2s+d+\frac{3}{2}}\left(t_f-t_i\right)$$

(25)

within the hypotheses that all parameters are independent of the time in the interval of integration, and with $K_{Drag}$ positive as the exponent $q = \left(2s+d+\frac{3}{2}\right)$ is usually negative.

The values of $r_c$, $c_c$ and $\rho_c$ are estimated as follows. The nebula's inner radius is taken as the reference distance $r_c$ and estimated by the proto-primary radius. The reference sound speed $c_c$ is found in the perfect gas approximation from a reference temperature, estimated from the proto-primary effective temperature $T_{eff}^*$

$$c_c = \sqrt{\frac{\gamma\mathcal{R}}{\mu}T_{eff}^*}$$

(26)

with $\mathcal{R}$ the perfect gas constant and $\mu$ the molecular mass, $\mu = 2.4\times10^{-3}$ kg/mole for a H/He gas mixture (Kusaka et al. 1970).

Assuming symmetries about the nebula rotation axis and about the midplane, the vertical hydrostatic equilibrium yields the specific mass to decrease exponentially with the altitude $z$ above the midplane

$$\rho(r,z) = \rho(r,0)\, e^{-\left(\frac{z}{h}\right)^2}$$

(27)

where $h$ is the scale height

$$h = \sqrt{\frac{2\, r^3}{\gamma\, GM^*}}\, c = \sqrt{\frac{2}{\gamma\, GM^*}\, \frac{r^{s+3}}{r_c^s}}\, c_c$$

(28)

neglecting the variation of the sound speed with the altitude in front of the specific mass variation. As $s$ is in the order of, or close to, -1, $h$ is approximately linear in $r$ and the nebula thickness is approximated by $2h \approx \frac{r}{10}$, corresponding to usual values in (28).

The mass $M_d$ of the nebular disc is written in function of $\rho_c$

$$M_d = 4\pi \frac{\rho_c}{r_c^d} \int_{r_c}^{r_e} \int_{0}^{h} e^{-\left(\frac{z}{h}\right)^2} r^{d+1}\, dz\, dr = \frac{\sqrt{\pi^3}}{10} \operatorname{erf}(1)\, H \rho_c r_c^3$$

(29)

where $r_e$ is the nebula external radius, erf the error function, and $H$ a function of the nebula radial extent $(\frac{r_e}{r_c})$ and $d$

$$H = H\left(\frac{r_e}{r_c}, d\right) = \begin{cases} \dfrac{\left(\frac{r_e}{r_c}\right)^{d+3} - 1}{d+3} & \text{if } d \neq -3 \\ \ln\left(\frac{r_e}{r_c}\right) & \text{if } d = -3 \end{cases}$$

(30)

Writing the initial nebula mass $M_d$ as a fraction $\mu$ of the primary mass $M^*$, $\rho_c$ is estimated from

$$\rho_c = \frac{10}{\sqrt{\pi^3} \operatorname{erf}(1)\, H}\, \mu\, M^* r_c^3$$

(31)

which can be replaced in $K_{Drag}$ (25) for the new variable $\mu$.

## 8. Evolution of $\frac{\bar{\beta}_i}{\bar{\beta}_f}$ for the four systems under the gas drag effect

The lower limit on the planetary nebula mass is estimated from the sum of the planets' present masses, increased up to Solar abundance (Cameron 1973), yielding

$$\mu_{min} = \left(\frac{M_d}{M_{Sun}}\right)_{min} \approx 3\%$$

The exponent $d$ of the initial gas specific mass distribution in the planetary nebula is usually taken between -1 and -3 for most models, e.g. -1.79 or -1.88 (Safronov 1969), -2.5 (Weidenschilling 1977b), -2.75 (Hayashi 1981), -2.72 (Coradini et al.1982), etc.

Neglecting all later dynamical effects, the final value of the mean distance ratio $\bar{\beta}_f$ at the nebula dissipation is taken to be the present value of $< \beta_f >$ between main regular secondaries.

Figure 1 shows for the planetary system the variation of the deviation $\Delta_{Drag}$ of $\frac{\bar{\beta}_i}{\bar{\beta}_f}$ from unity in function of $d$ for $-4 \leq d \leq -1$, $s = -1$ and $s = -\frac{1}{2}$, with, first, $(t_f - t_i) = 10^6$ and $10^7$ yrs for $\mu = 5\%$ (curves A to D), and, second, $(t_f - t_i) = 10^4$ yrs for $\mu = 1$ (curves E and F; disc of one Solar mass, limited to Neptune's actual orbit).

We find that, first, $\Delta_{Drag}$ is always negative, which means that, if only gas drag is considered, the initial mean distance ratio in the planetary nebula was smaller; second, $\Delta_{Drag}$ tends toward 0 for small negative values of $d$ and pass by minima (not shown on the figure) for smaller negative values of $d$; third, in the 5% Solar mass nebula case, $\Delta_{Drag}$ stays small for $d \geq -3$, respectively less than 4 % and 11% for $(t_f - t_i) = 10^6$ and $10^7$ yrs; fourth, $\Delta_{Drag}$ increases when $(t_f - t_i)$ is increased or when $s$ is decreased, for other parameters constant; and lastly, in the one Solar mass nebula case, $\Delta_{Drag}$ is surprisingly less than 1% for $d \geq -3$ and for $(t_f - t_i)$ =$10^4$ yrs.

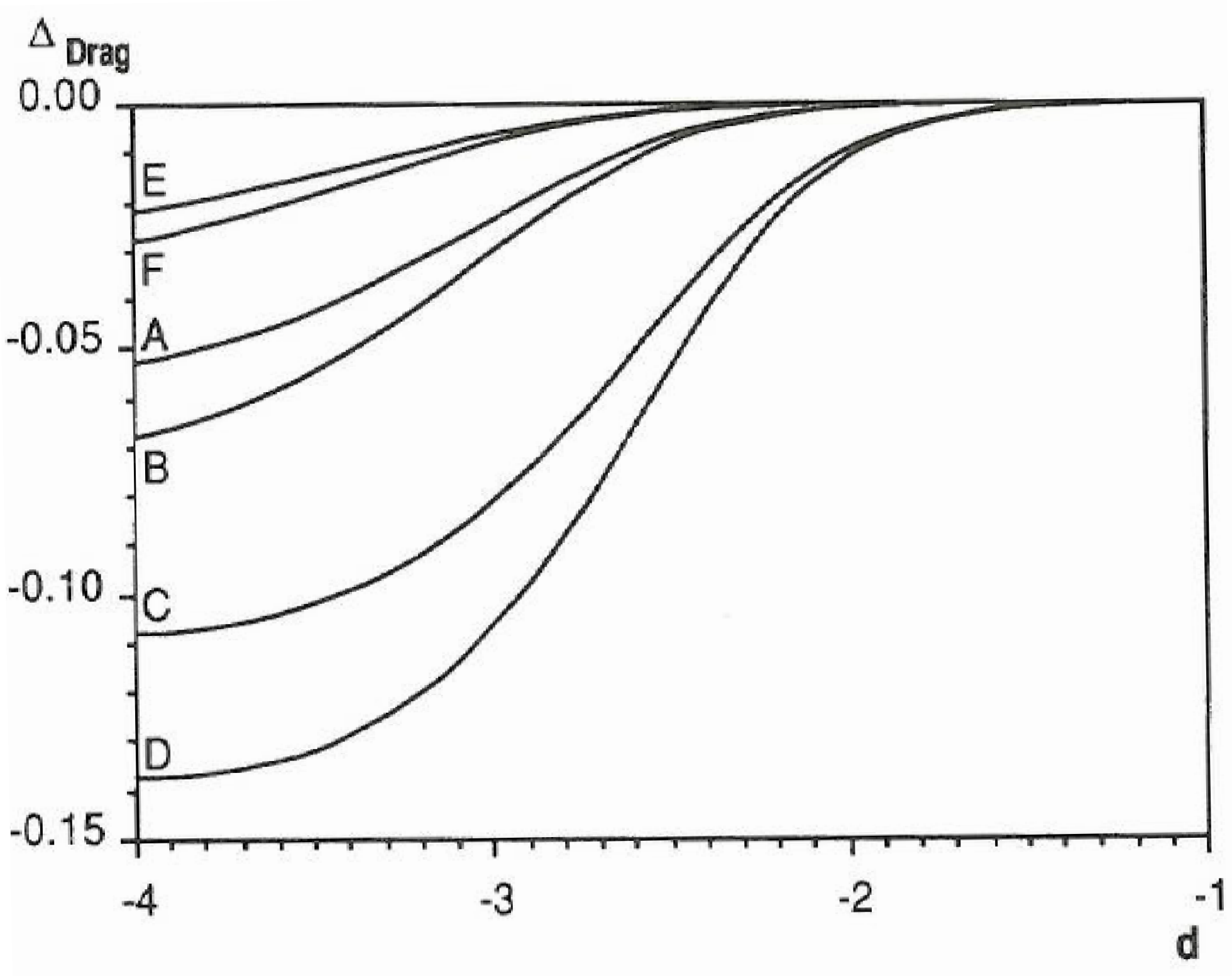

Figure 1 : Planetary nebula, deviation $\Delta_{Drag}$ from unity of the ratio of initial to final mean distance ratios $\frac{\bar{\beta}_i}{\bar{\beta}_f}$ in function of exponents $d$ of the initial gas specific mass power-law distribution for $n$ = 8 (Mercury to Neptune); low nebula mass case $\mu = \frac{M_d}{M_{Sun}} = 5\%$; $(t_f - t_i) = 10^6$ yrs for $s = -1$ (curve A) and $s = -\frac{1}{2}$ (B), $(t_f - t_i) = 10^7$ yrs for $s = -1$ (C) and $s = -\frac{1}{2}$ (D); high nebula mass case $\mu = 1$, $(t_f - t_i) = 10^4$ for $s = -1$ (E) and $s = -\frac{1}{2}$ (F). Other parameters in (25) to (31) in text are $r_c = 3.6\times10^{10}$ m, $r_e = 5.3\times10^{12}$ m, $T^*_{eff} = 3.1\times10^3$ °K, $c_c = 4\times10^3$ m/s, $\gamma = \frac{3}{2}$.

For the satellite nebulae, lower limits on masses are estimated from the sum of the satellites' increased masses in Solar (or central planet's) composition, yielding $\mu_{min}$ about 1.5% for the three nebulae. However, this method provides only first-guessed values on lower limits, as the satellite nebulae do not necessarily need to be of solar or central planet compositions, e.g. the nebular material could have been depleted in heavy elements by an early primary core accretion or augmented by later capture of heliocentric material (Weidenschilling 1982). The exponent $d$ in the satellite nebulae can be estimated by power-law fitting of satellites' increased masses, spread over annular rings centred on present orbits, yielding values of $d$ between -2 and -3 for the Jovian, Saturnian and Uranian systems, although with some reserves for the Saturnian system (see e.g. Pollack and Consolmagno 1984). The scale height used in the estimation of $\rho_c$ via $M_d$ is sometimes larger in some models than the value $r/10$ and varies in different zones of the nebulae, with typical values about $r/4$ to $r/8$ (Harris 1984; Safronov et al. 1986). As these scale heights lower the values of $\rho_c$, reducing the drag effect, we keep the standard value $r/10$. Furthermore, if one accounts only for the lower gas specific mass in tidally cleared tunnels, the corresponding specific mass at satellites' distances could be grossly given by power-law distributions (19) with similar exponents $d$ as in initial nebulae, but with lower reference values $\rho_c$, for which corresponding values of $\mu$ less than $\mu_{min}$ may be relevant.

In Figures 2 to 4, the variations $\Delta_{Drag}$ in the Jovian, Saturnian and Uranian nebulae are given in function of the nebula/planet mass fraction $\mu$, $10^{-4} < \mu < 10^{-1}$, for $d$ = -2 and -2.5 and for $(t_f - t_i)$ = $10^3$, $10^4$ and $10^5$ yrs (and $10^6$ yrs in the Jovian case), considering optically thick nebulae with $s$ = -1.

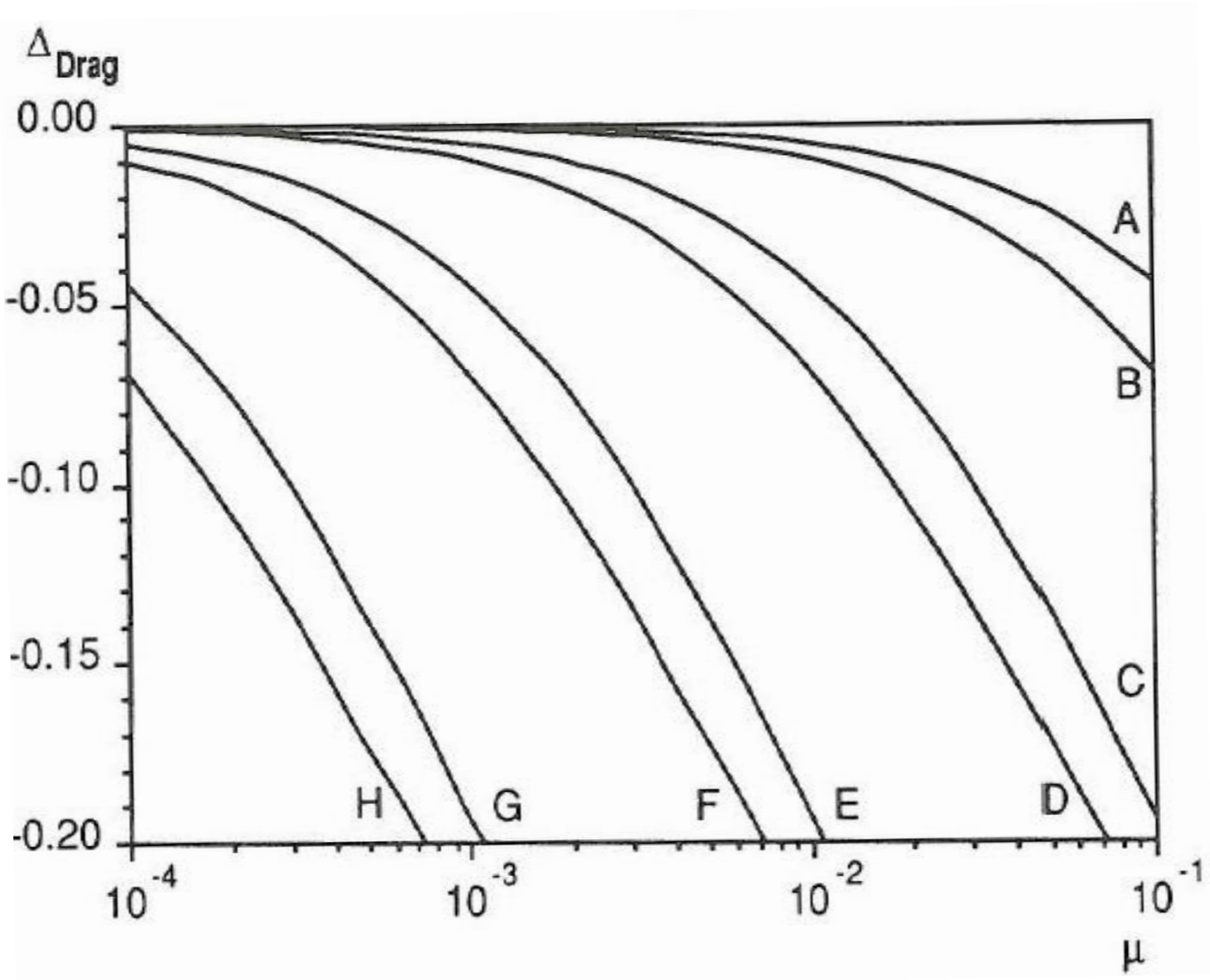

Figure 2 : Jovian nebula, deviation $\Delta_{Drag}$ vs $\mu$; $n$ = 3 (Io to Callisto): $(t_f - t_i)$ = $10^3$ yrs for $d$ = -2 (curve A) and -2.5 (B), $(t_f - t_i)$ = $10^4$ yrs for $d$ = -2 (C) and -2.5 (D), $(t_f - t_i)$ = $10^5$ yrs for $d$ = -2 (E) and -2.5 (F), $(t_f - t_i)$ = $10^6$ yrs for $d$ = -2 (G) and -2.5 (H). Other parameters are $r_c$ = 9.1×10$^7$ m, $r_e$ = 2.2×10$^9$ m, $T^*_{eff}$ = 600 °K, $c_c$ = 1.8×10$^3$ m/s, $s$ = -1, $\gamma = \frac{3}{2}$.

We find that $\Delta_{Drag}$ is always negative and that it increases with $(t_f - t_i)$, or with $\mu$, or for decreasingly negative values of $d$.

For the Jovian system (lo to Callisto), $\Delta_{Drag}$ is less than about 5% for $(t_f - t_i)$ = $10^4$ yrs if $\mu$ is less than about 1%, and for $(t_f - t_i)$ = $10^5$ yrs if $\mu$ is less than about $10^{-3}$.

For the Saturnian system, the two series of curves for the sets Mimas to Iapetus and Mimas to Rhea diverge for increasing $\mu$ and $(t_f - t_i)$. $\Delta_{Drag}$ is less than about 5% for $(t_f - t_i)$ = $10^4$ yrs if $\mu$ is less than about $5\times10^{-4}$. Note that we took values of the initial proto-Saturn radius and effective temperature smaller than in Bodenheimer's (et al. 1980) model to account for the presence of Mimas and for a cooler nebula (Pollack 1985) and that we considered a Saturnian nebula extending to Iapetus' actual orbit, which makes it more than one and half more extended than Jupiter's nebula.

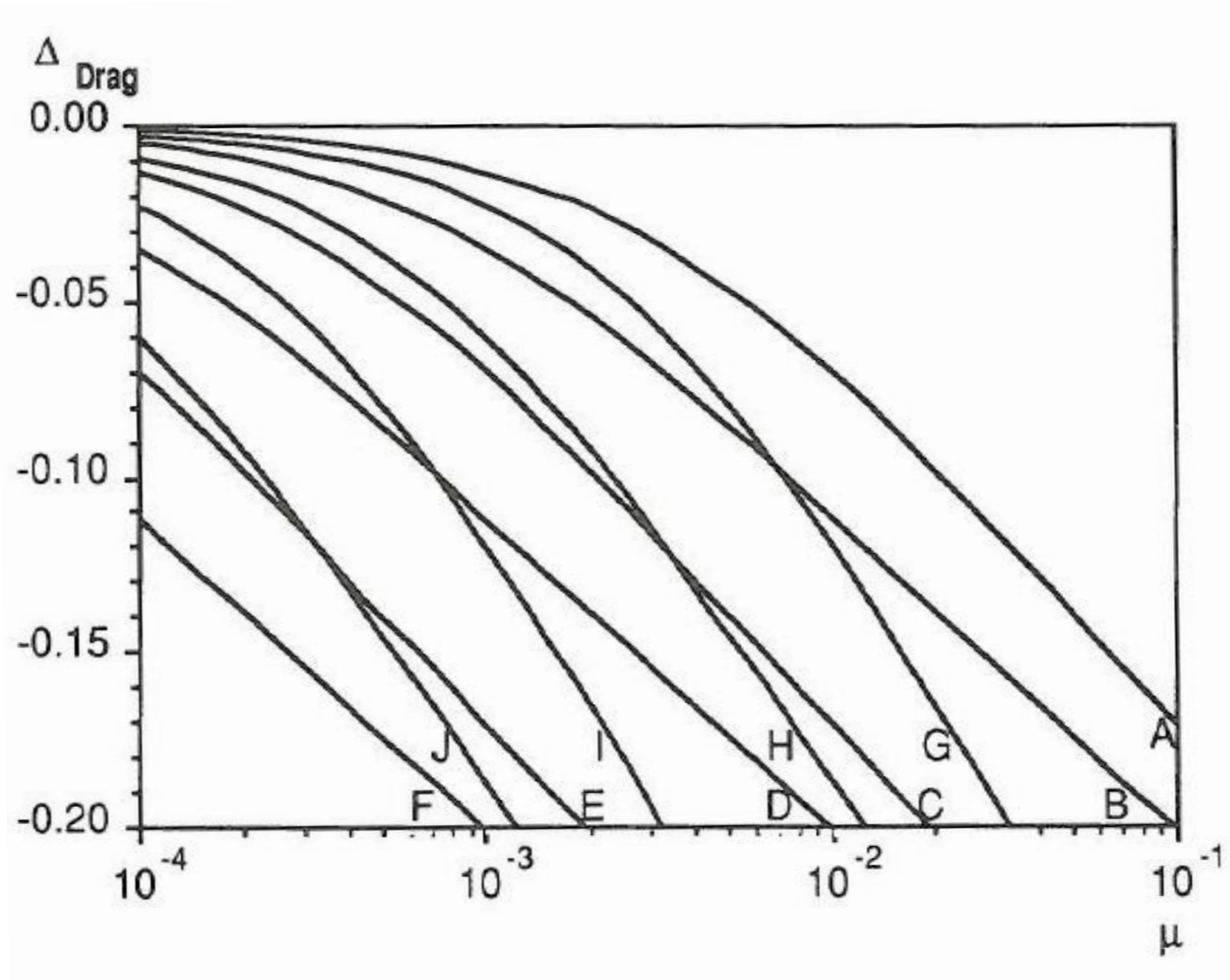

Figure 3 : Saturnian nebula, deviation $\Delta_{Drag}$ vs $\mu$; first case $n$ = 7 (Mimas to Iapetus): $(t_f - t_i)$ = $10^3$ yrs for $d$ = -2 (curve A) and -2.5 (B), $(t_f - t_i)$ = $10^4$ yrs for $d$ = -2 (C) and -2.5 (D), $(t_f - t_i)$ = $10^5$ yrs for $d$ = -2 (E) and -2.5 (F); second case $n$ = 4 (Mimas to Rhea): $(t_f - t_i)$ = $10^3$ yrs for $d$ = -2 (G) and -2.5 (H), $(t_f - t_i)$ = $10^4$ yrs for $d$ = -2 (I) and -2.5 (J). Other parameters are $r_c$ = $1.5\times10^8$ m, $r_e$ = $3.6\times10^9$ m, $T^*_{eff}$ = 250 °K, $c_c$ = $10^3$ m/s, $s$ = -1, $\gamma = \frac{3}{2}$.

For the Uranian system (Ariel to Oberon, although similar curves are found for the set Miranda to Oberon), $\Delta_{Drag}$ is less than about 5% for $(t_f - t_i)$ = $10^4$ yrs if $\mu$ is less than about $2\times10^{-3}$, and for $(t_f - t_i) = 10^5$ if $\mu$ is less than about $2\times10^{-4}$, assuming Uranus' initial radius is twice its present value.

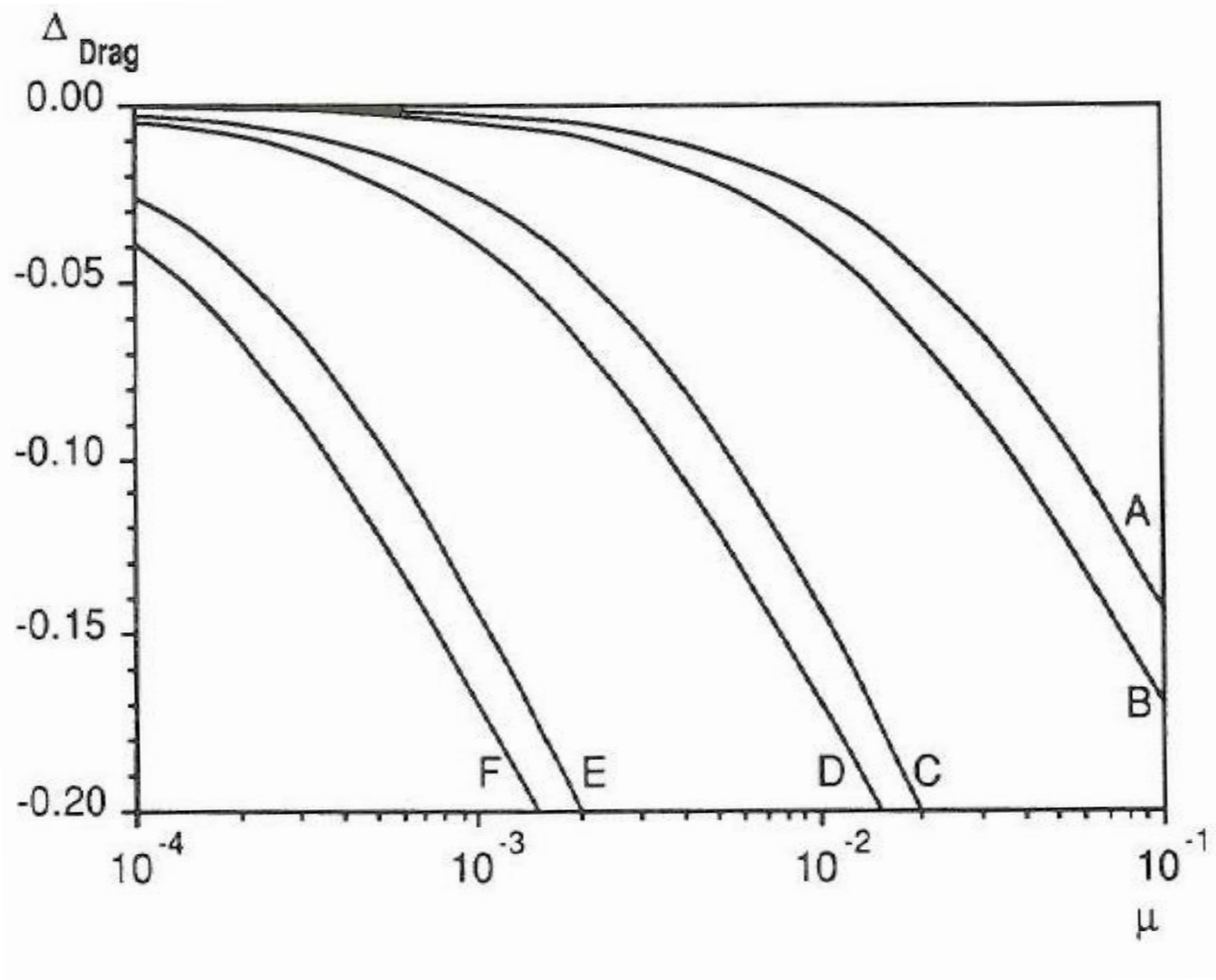

Figure 4 : Uranian nebula, deviation $\Delta_{Drag}$ vs $\mu$; $n$ = 3 (Ariel to Oberon): $(t_f - t_i) = 10^3$ yrs for $d$ = -2 (curve A) and -2.5 (B), $(t_f - t_i) = 10^4$ yrs for $d$ = -2 (C) and -2.5 (D), $(t_f - t_i) = 10^5$ yrs for $d$ = -2 (E) and -2.5 (F). Other parameters are $r_c = 5\times10^7$ m (assumed), $r_e = 7\times10^8$ m, $T^*_{eff}$ = 100 °K, $c_c = 7.2\times10^2$ m/s, $s$ = -1, $\gamma = \frac{3}{2}$.

In all cases, $\Delta_{Drag}$ is reduced for shorter periods $(t_f - t_i)$, which should be considered if the tunnelling effect reduces the effective drag period, and also for smaller values of the initial nebula relative mass $\mu$, if one accounts for the possibly less dense gas in tidally cleared tunnels. Combining the effects of orbital resonance and gas drag, the semi-major axis of a body in resonance evolving in an eccentricity-reducing medium would secularly increase or decrease if its orbit is outside or inside the perturber orbit (Greenberg 1978). However, stable capture in resonance under gas drag is only possible for first order ($\frac{j}{j+1}$) exterior resonances ($j$ integer), i.e., the perturbed body's orbit is exterior to the perturber one (Weidenschilling and Davis 1985). As already mentioned, resonances would freeze the distance ratio of the two bodies and reduce the effect of gas drag on the evolution of the mean distance ratios.

**9. Discussion and implications for planetary system formation**

It was shown that the orbital evolutions of planets and Jovian, Saturnian and Uranian satellites, due to primaries' tides only and after the initial nebulae dissipations, are such as to reduce the mean distance ratios among the main regular secondaries, although in a very limited way, and the mean distance ratios can be considered as approximately conserved. However, the results for the satellite systems depend on the adopted values of $Q$. The resonances between satellites do not affect these results, as local freezing of distances between resonant satellites reinforces the approximate conservation of the mean distance ratios. We showed further that the mean distance ratios is increased by the nebular gas drag effect after the secondaries accretion but, for usually assumed values of physical characteristics in initial nebulae, this increase is small for the planetary high mass model, but it may also be small in the planetary low mass model for $d \geq -3$ and in the giant planet satellite systems, depending on the nebulae lifetimes and on the periocl of effective gas drag due to the tunnelling effect.

However, these results must be interpreted with caution in view of the hypotheses and assumptions. The results of the approximate conservation of the planetary system mean distance ratio over a period of $4.5\times10^9$ years, considering only the primary tides effect, is in contradiction with the results of some numerical integration (Laskar 1989; Tsiganis et al. 2005; Desch 2007) showing that past orbital evolution of planets may have been chaotic, although the extent and implications of this chaotic behaviour remain an active area of research.

In modern planet formation models, orbital architectures are shaped by a combination of accretion processes, disk-driven migration, and dynamical interactions. Migration induced by disk–planet interactions (Lin and Papaloizou,1986; Goldreich and Tremaine 1980; Kley and Nelson 2012) can lead to substantial radial displacement of individual bodies. However, such migration often occurs in a coordinated manner, particularly in systems where planets are trapped in resonant chains (Batygin 2015; Batygin and Morbidelli 2026), which can preserve relative spacing properties.

Recent studies suggest that planetary systems may emerge from rings of planetesimals or pressure maxima in disks, leading naturally to characteristic spacing scales (Izidoro et al. 2022). Similarly, pebble accretion and radial drift processes can generate diversity in planetary architectures while maintaining underlying structural regularities (McCloat et al. 2025). In this context, the approximate conservation of mean distance ratios found in the present work is consistent with the idea that large-scale spacing properties are established early and only weakly modified thereafter.

The diversity of exoplanetary systems further supports this interpretation. While systems may undergo dynamical instabilities or planet loss (Raymond et al. 2020), surviving systems often retain coherent spacing patterns. This "survivor bias" implies that observed architectures may preferentially reflect dynamically stable configurations that preserve initial spacing characteristics.

Finally, the analytical framework developed here complements numerical studies of planetary system formation (Ida and Lin 2004a, b; Clement et al. 2024) by providing a simple global diagnostic of system evolution. The mean distance ratio may therefore serve as a useful parameter for comparing different formation scenarios and for interpreting the growing body of observational data on exoplanetary systems.

## 10. Conclusions

In this work, we have investigated the evolution of mean distance ratios in planetary and satellite systems under the influence of primary tidal dissipation and nebular gas drag. Using a general analytical framework, we have shown that both processes produce only limited variations of the mean distance ratio over characteristic evolutionary timescales.

For tidal evolution acting over the age of the Solar System, the resulting deviations are found to be negligible in all considered systems, confirming that tidal dissipation alone cannot significantly alter the global spacing structure. Similarly, for a wide range of nebular models and plausible disk lifetimes, gas drag induces only moderate changes in mean distance ratios, particularly when accounting for reduced effective drag due to disk depletion or gap formation. These results suggest that the present-day mean distance ratios of planetary and regular satellite systems largely preserve information about their primordial configurations. In this context, the approximate geometric spacing observed in these systems may be interpreted as a fossil signature of the early stages of disk evolution and fragmentation.

This conclusion gains additional relevance in light of recent observations of compact exoplanetary systems, which frequently exhibit regular orbital spacing and near-resonant configurations. The approximate conservation of mean distance ratios may therefore provide a useful diagnostic for distinguishing between competing formation and migration scenarios, and for constraining the physical conditions in protoplanetary disks.

Future work should aim to extend the present analytical approach by incorporating modern numerical simulations of planet–disk interactions and multi-body dynamics, as well as by applying the framework to observed exoplanetary systems. Such developments would help clarify the extent to which orbital spacing relations can serve as robust tracers of planetary system formation processes.

## Acknowledgement

We thank Pr. P. Pâquet, University of Louvain, and Pr. V. Dehant, Royal Observatory of Belgium, for discussions on the values of $k_2$ and $Q$.

## References

Barnes, R., Deitrick, R., Greenberg, R., Quinn, T.R., Raymond, S.N. (2015). Long-lived chaotic orbital evolution of exoplanets in mean motion resonances with mutual inclinations. *Astrophys. J.*, 801, 101. https://doi.org/10.1088/0004-637X/801/2/101

Batygin, K. (2015). Capture of planets into mean-motion resonances and the origins of extrasolar orbital architectures. *Mon. Not. R. Astron. Soc.*, 451, 2589–2609. https://doi.org/10.1093/mnras/stv1063

Batygin, K., Morbidelli, A. (2026). Origins of compact mean-motion resonances: Evidence for long-range migration and the case of Kepler-36. *arXiv preprint*, arXiv:2603.27093

Bodenheimer, P., Grossman, A.S., De Campli, W.M., Marcy, G., Pollack, J.B. (1980). Calculations of the evolution of the giant planets. *Icarus*, 41, 293–308

Bourrier, V., Attia, M., Mallonn, M., Marret, A., Lendl, M., et al. (2023). DREAM: I. Orbital architecture orrery. *Astron. Astrophys.*, 669, A63. https://doi.org/10.1051/0004-6361/202245004

Burns, J.A. (1976). An elementary derivation of the perturbation equations of celestial mechanics. *Am. J. Phys.*, 44, 944–949

Burns, J.A. (1977). Orbital evolution. In: Burns, J.A. (ed.) *Planetary Satellites*, pp. 113–156. Univ. Arizona Press

Burns, J.A. (1982). The dynamical evolution of the Solar System. In: Brahic, A. (ed.) *Formation of Planetary Systems*, CNES, Cepadues Editions, Toulouse, France, pp. 403–501

Burns, J.A., Showalter, M.R., Morfill, G.E. (1984). The ethereal rings of Jupiter and Saturn. In: Greenberg, R., Brahic, A. (eds) *Planetary Rings*, Univ. of Arizona Press, Tucson, pp. 200–272

Cameron, A.G.W. (1973). Abundances of the elements in the Solar System. *Space Sci. Rev.*, 15, 121–146

Cameron, A.G.W., Pine, M.R. (1973). Numerical models of the primitive solar nebula. *Icarus*, 18, 377–406

Clement, M.S., Izidoro, A., Raymond, S.N., Deienno, R. (2024). Formation of terrestrial planets. *arXiv preprint*, arXiv:2411.03453

Coradini, A., Federico, C., Magni, G. (1981). Gravitational instabilities in satellite disks and formation of regular satellites. *Astron. Astrophys.*, 99, 255–261

Coradini, A., Federico, C., Magni, G. (1982). Some remarks on the formation of terrestrial planets. In: Coradini, A., Fulchignoni, M. (eds) *Comparative Study of the Planets*, D. Reidel Publ. Co., Dordrecht, The Netherlands, pp. 3–24

Cox, J.P., Giuli, R.T. (1968). *Principles of Stellar Structure, Vol.2 Applications to Stars*, Gordon and Breach, Science Publishers, New York, pp. 944-1028.

Desch, S.J. (2007). Mass distribution and planet formation in the Solar nebula. *Astrophys. J.*, 671, 878–893. https://doi.org/10.1086/522825

Ezer, D., Cameron, A.G.W. (1965). A study of solar evolution. *Can. J. Phys.*, 43, 1497–1517

Fabrycky, D.C., Lissauer, J.J., Ragozzine, D., Rowe, J.F., Steffen, J.H., et al. (2014). Architecture of Kepler's multi-transiting systems II, New investigations with twice as many candidates. *Astrophys. J.*, 790, 146. https://doi.org/10.1088/0004-637X/790/2/146

Gavrilov, S.V., Zharkov, V.N. (1977). Love numbers of the giant planets. *Icarus*, 32, 443–449

Goldreich, P. (1965). An explanation of the frequent occurrence of commensurable motions. *Mon. Not. R. Astron. Soc.*, 130, 159–181

Goldreich, P., Soter, S. (1966). Q in the Solar System. *Icarus*, 5, 375–389

Goldreich, P., Tremaine, S. (1980). Disk-satellite interactions. *Astrophys. J.*, 241, 425–441. https://doi.org/10.1086/158356

Grasboke, H.C., Pollack, J.B., Grossman, A.S., Olness, R.J. (1975). The structure and evolution of Jupiter: The fluid contraction stage. *Astrophys. J.*, 199, 265–281

Greenberg, R. (1978). Orbital resonance in a dissipative medium. *Icarus*, 33, 62–73

Greenberg, R. (1982). Planetesimals to planets. In: Brahic, A. (ed.) *Formation of Planetary Systems*, CNES, Cepadues Editions, Toulouse, France, pp. 515–569

Greenberg, R. (1987). Galilean satellites: evolutionary paths in deep resonance. *Icarus*, 70, 334–347

Greenberg, R., Goldstein, S.J., Jacobs, K.C. (1986). Orbital acceleration and the energy budget in the Galilean satellite system. *Nature*, 323, 789–791

Harris, A.W. (1978). Satellite formation II. *Icarus*, 34, 128–145

Harris, A.W. (1984). The origin and evolution of planetary rings. In: Greenberg, R., Brahic, A. (eds) *Planetary Rings*, Univ. of Arizona Press, Tucson, pp. 641-659.

Hayashi, C. (1981). Structure of the solar nebula, growth and decay of magnetic fields and effects of magnetic and turbulent viscosities on the nebula. *Prog. Theor. Phys. Suppl.*, 70, 35–53

Hayashi, C., Nakazawa, K., Nakagawa, Y. (1985). Formation of the solar system. In Black D., Matthews M. (eds) *Protostars and Planets II*. Univ. of Arizona Press, Tucson, pp. 1100-1153

Henrard, J. (1983). Orbital evolution of the Galilean satellites: Capture into resonance. *Icarus*, 53, 55–67

Iben, I. (1965). Stellar evolution I: The approach to the main sequence. *Astrophys. J.*, 141, 993–1018

Ida, S., Lin, D.N.C. (2004a). Toward a deterministic model of planetary formation. I. A desert in the mass and semimajor axis distributions of extrasolar planets. *Astrophys. J.*, 604, 388–413. https://doi.org/10.1086/381724

Ida, S., Lin, D.N.C. (2004b). Toward a deterministic model of planetary formation. II. The formation and retention of gas giant planets around stars with a range of metallicities. *Astrophys. J.*, 616. https://doi.org/10.1086/424830

Izidoro, A., Dasgupta, R., Raymond, S.N., Deienno, R., Bitsch, B., et al. (2022). Planetesimal rings as the cause of the Solar System's planetary architecture. *Nat. Astron.*, 6, 357–366. https://doi.org/10.1038/s41550-021-01557-z

Kley, W., Nelson, R.P. (2012). Planet-Disk Interaction and Orbital Evolution. *Annu. Rev. Astron. Astrophys.*, 50, 211–249

Kusaka, T., Nakano, T., Hayashi, C. (1970). Growth of solid particles in the primordial solar nebula. *Prog. Theor. Phys.*, 44, 1580–1596

Laskar, J. (1989). A numerical experiment on the chaotic behaviour of the Solar System. *Nature*, 338, 237–238

Lecar, M. (1973). Bode's law. *Nature*, 242, 318–319

Lin, D.N.C. (1981). Convective accretion disk model for the primitive solar nebula. *Astrophys. J.*, 246, 972–984

Lin, D.N.C., Papaloizou, J. (1979). Tidal torques on accretion disks in binary systems with extreme mass ratios. *Mon. Not. R. Astron. Soc.*, 186, 799–812

Lin, D.N.C., Papaloizou, J. (1980). On the structure and evolution of the primordial solar nebula. *Mon. Not. R. Astron. Soc.*, 191, 37–48

Lin, D.N.C., Papaloizou, J. (1986). On the Tidal Interaction between Protoplanets and the Protoplanetary Disk. III. Orbital Migration of Protoplanets. *Astrophys. J.*, 309, 846. https://doi.org/10.1086/164653

Lissauer, J.J., Ragozzine, D., Fabrycky, D.C., Steffen, J.H., Ford, E.B., et al. (2011). Architecture and dynamics of Kepler's candidate multiple transiting planet systems. *Astrophys. J. Suppl.*, 197, 8

Lynden-Bell, D., Pringle, J.E. (1974). The evolution of viscous disks and the origin of nebular variables. *Mon. Not. R. Astron. Soc.*, 168, 603–637

McCloat, S., Mulders, G.D., Fieber-Beyer, S. (2025). Diversity in planetary architectures from pebble accretion: Water delivery to the habitable zone with pebble snow. *arXiv preprint*, arXiv:2509.14101

Morbidelli, A. (2013). Dynamical Evolution of Planetary Systems. In: Oswalt, T.D., French, L.M., Kalas, P. (eds) *Planets, Stars and Stellar Systems*. Springer, Dordrecht. doi: 10.1007/978-94-007-5606-9_2.

Morbidelli, A., Tsiganis, K., Crida, A., Levison, H.F., Gomes, R. (2007). Dynamics of the giant planets of the Solar System in the gaseous proto-planetary disk and relationship to the current orbital architecture. *Astron. J.*, 134, 1790–1798

Morbidelli, A., Lunine, J.I., O'Brien, D.P., Raymond, S.N., Walsh, K.J. (2012). Building Terrestrial Planets. *Annu. Rev. Earth Planet. Sci.*, 40, 251–275

Peale, S.J. (1986). Orbital resonances, unusual configurations & exotic rotation states among planetary satellites. In Burns J.A., Matthews, M.S. (eds) *Satellites*. Univ. of Arizona Press, Tucson, pp.159-223

Pletser, V. (1986). Exponential distance laws for satellite systems. *Earth Moon Planets*, 36, 193–210

Pletser, V. (1987). Spacing of accretion site locations in random planetary-like systems. In: *Proc. Comparative Planetology Workshop*, CNES, Toulouse (Available at https://www.researchgate.net/publication/257880590_Spacing_of_accretion_site_locations_in_random_planetary-like_systems)

Pletser, V. (1988a). Revised exponential distance relation for the Uranian system after the Voyager 2 fly-by. *Earth Moon Planets*, 41, 295–300

Pletser, V. (1988b). Exponential distance relations in planetary-like systems generated at random. *Earth Moon Planets*, 42, 1–18

Pletser, V. (1990). *On exponential distance relations in planetary and satellite systems, observations and origin*. PhD Thesis, Catholic University of Louvain (Available at

https://www.researchgate.net/publication/257927392_On_exponential_distance_relations_in_planetary_and_satellite_systems_observations_and_origin)

Pletser, V. (2017a). Non-randomness of exponential distance relations in the planetary system: an answer to Lecar. *Adv. Space Res.*, 60, 2314–2318

Pletser, V. (2017b). Lecar's visual comparison method to assess the randomness of Bode's law: an answer. *arXiv preprint*, arXiv:1709.02704

Pletser, V., Basano, L. (2017). Exponential distance relation and near resonances in the TRAPPIST-1 planetary system. *arXiv preprint*, arXiv:1703.04545

Pollack, J.B. (1985). Formation of the giant planets and their satellite-ring systems: an overview. In Black D., Matthews M. (eds.) *Protostars and Planets II.* Univ. of Arizona Press, Tucson, pp. 791-831.

Pollack, J.B., Grossman, A.S., Moore, R., Grasboke, H.C. (1977). A calculation of Saturn's gravitational contraction history. *Icarus*, 30, 111–128

Pollack, J.B., Consolmagno, G. (1984). Origin and evolution of the Saturn system. In GehreIs T., Matthews M. (eds.) *Saturn*. Univ. of Arizona Press, Tucson, pp. 811-866.

Raymond, S.N., Boulet, T., Izidoro, A., Esteves, L., Bitsch, B. (2018). Migration-driven diversity of super-Earth compositions. *Mon. Not. R. Astron. Soc. Lett.*, 479, L81–L85, doi 10.1093/mnrasl/sly100.

Raymond, S.N., Izidoro, A., Morbidelli, A. (2018). Solar System formation in the context of exoplanets. *arXiv preprint*, arXiv:1812.01033

Raymond, S.N., Kaib, N.A., Armitage, P.J., Fortney, J.J. (2020). Survivor Bias: Divergent fates of the Solar System's ejected versus persisting planetesimals. *Astrophys. J. Lett.*, 904, L4. https://doi.org/10.3847/2041-8213/abc55f

Raymond, S.N., Izidoro, A., Morbidelli, A. (2020). Solar System formation in the context of extra-solar planets. In Meadows V.S., Arney G.N., Schmidt B.E., Des Marais D.J. (eds) *Planetary Astrobiology*, Space Science Series, University of Arizona Press. doi 10.2458/azu_uapress_9780816540068.

Raymond, S.N., Izidoro, A., Morbidelli, A. (2022). Solar System formation in the context of exoplanets, *Exoplanets in Our Backyard 2022*. LPI Contrib. No. 2687. Available online at https://www.hou.usra.edu/meetings/exoplanets2022/pdf/3018.pdf

Safronov, V.S. (1969). *Evolution of the Protoplanetary Cloud and the Formation of the Earth and Planets*. Nauka Press, Moscow, transl. NASA TTF-677 (1972)

Safronov, V.S., Pechernikova, G.V., Ruskol, E.L., Vitjazev, A.V. (1986). Protosatellite swarms. In Burns J.A., Matthews M.S. (eds.) *Satellites*. Univ. of Arizona Press, Tucson, pp. 89-116

Schubert, G., Spohn, T., Reynolds, R.T. (1986). Thermal histories, compositions and internal structures of the moons of the solar system. In Burns J.A., Matthews M.S. (eds.) *Satellites*. Univ. of Arizona Press, Tucson, pp. 224-292

Showalter, M.R., Burns, J.A., Cuzzi, J.N., Pollack, J.B. (1985). Discovery of Jupiter's 'gossamer' ring. *Nature*, 316, 526–528

Sinclair, A.T. (1975). The orbital resonance amongst the Galilean satellites of Jupiter. *Mon. Not. R. Astron. Soc.*, 171, 59–72

Stevenson, D.J., Harris, A.W., Lunine, J.I. (1986). Origins of satellites. In Burns J.A., Matthews M.S. (eds.) *Satellites*. Univ. of Arizona Press, Tucson, pp. 39-88.

Tsiganis, K., Gomes, R., Morbidelli, A., Levison, H.F. (2005). Origin of the orbital architecture of the giant planets of the Solar System. *Nature Letters*, 435, 459–461. https://doi.org/10.1038/nature03539

Weidenschilling, S.J. (1977a). Aerodynamics of solid bodies in the solar nebula. *Mon. Not. R. Astron. Soc.*, 180, 57–70

Weidenschilling, S.J. (1977b). The distribution of mass in the planetary system and solar nebula. *Astrophys. Space Sci.*, 51, 153–158

Weidenschilling, S.J. (1982). Origin of regular satellites. In Coradini A., Fulchignoni M. (eds.) *The Comparative Study of the Planets*. D. Reidel Publ. Co, Dordrecht, The Netherlands, pp. 49-59

Weidenschilling, S.J., Davis, D.R. (1985). Orbital resonances in the solar nebula: Implications for planetary accretion. *Icarus*, 62, 16–29

Whipple, F.L. (1972). On certain aerodynamic processes for asteroids and comets. In Elvius A. (ed.) *From Plasma to Planets*. Wiley, New York, pp. 211-232.

Yoder, C.F. (1979). How tidal heating in Io drives the Galilean orbital resonance locks. *Nature*, 279, 767–770

Yoder, C.F., Peale, S.J. (1981). Tides of Io. *Icarus*, 47, 1–35

Youdin, A.N., Shu, F.H. (2002). Planetesimal formation by gravitational instability. *Astrophys. J.*, 580, 494–505. https://doi.org/10.1086/343109

Youdin, A.N., Kenyon, S.J. (2013). From Disks to Planets. In Oswalt T.D., French L.M., Kalas P. (eds) *Planets, Stars and Stellar Systems*. Springer, Dordrecht. doi: 10.1007/978-94-007-5606-9_1.